%% file: main.tex
\documentclass[nopreprintline,5p,11pt,authoryear]{elsarticle}

\input{Packages}
\usepackage{float}

\begin{document}

\begin{frontmatter}

\title{Spacecraft Angular Rate Estimation via Event-Based Camera Sensing}

\author[inst1]{V. Franzese\corref{cor1}}
\ead{vittorio.franzese@uni.lu}
\cortext[cor1]{Corresponding author}

\author[inst1]{M. El Hariry}
\ead{matteo.elhariry@uni.lu}

\affiliation[inst1]{organisation={Interdisciplinary Centre for Security, Reliability and Trust (SnT), University of Luxembourg},
            addressline={29 Av. J. F. Kennedy}, 
            city={Luxembourg},
            postcode={1855},
            country={Luxembourg}}
            
\begin{abstract}
This paper presents a method for determining spacecraft angular rates using event-based camera sensing. This is achieved by analyzing the temporal distribution of brightness events triggered by the apparent motion of stars. The location and polarity of the events are used to infer the apparent motion field of the stars, which is, in turn, employed to estimate the observer angular velocity in the camera frame. This can be converted to the spacecraft angular rates provided an attitude reference. The method is validated through numerical simulation for a synthetic dataset of event streams generated on random spacecraft pointing and rates conditions. The accuracy of the method is assessed, demonstrating its potential to complement or replace conventional rate sensors in spacecraft systems using event camera sensing.
\end{abstract}

\begin{keyword}
Angular Rate Estimation \sep Event-based Camera \sep Spacecraft Sensor \sep Event Sensing
\end{keyword}

\end{frontmatter}

\nolinenumbers

\section{Introduction} \label{sec:introduction}

Knowledge of a spacecraft angular rate is essential to accomplish several tasks, such as attitude control and navigation, thus meeting requirements and objectives of space missions \citep{crassidis2007survey}. Traditionally, this information is provided by gyroscopes, which deliver measurements of angular velocity. While widely used, gyroscopes suffer from drift over time, can degrade under radiation or mechanical stress, and may fail \citep{venkateswaran2002analytical}. In case of failure and lack of redundancy of such sensors, a spacecraft may lose its ability to operate nominally, potentially jeopardising the success of the mission. Therefore, it is beneficial to consider alternative methods and sensors to complement angular rate estimation for spacecraft. Alternative techniques have been explored so far to this aim. These are based on star trackers and sun sensors, where angular velocity is inferred from successive attitude measurements \citep{jo2015optimal}, and also optical flow methods using frame-based cameras, leveraging the apparent motion of celestial features on the focal plane of the camera \citep{liebe2004toward}. 

{Event-based sensors, inspired by neuromorphic vision, offer a new sensing approach with respect to traditional ones \citep{gehrig2024low}. Unlike conventional frame-based sensors, event cameras asynchronously report changes in brightness at each pixel with microsecond-level temporal resolution and low latency \citep{rebecq2019high, elms2024event}. This provides sparse but information-rich data, with advantages in dynamic scenarios, low-light conditions, and power-constrained systems, as is the case for satellites in space \citep{izzo2023neuromorphic}. This is particularly true for small satellites \citep{di2022erc}. Event-based sensors have been used for angular rate estimation for Earth applications \citep{gallego2017accurate, shiba2024secrets}. Also, event cameras started being investigated for use in space engineering owing to their peculiarities and for the new sensing approach they offer. Examples include the use of event sensing to update the location of stars in images within a Kalman filter formulation \citep{ng2022asynchronous}, or to estimate a spacecraft attitude \citep{chin2019star, bagchi2020event}.} 

{While different works have focused on attitude estimation using event-based star sensing, this paper details a method for determining spacecraft angular rates using event-based camera sensing and relying on unidentified stars. This is also achieved through information fusion by an orthogonal camera setup to increase the angular rate estimation accuracy.} By analyzing the temporal distribution of brightness events triggered by the apparent motion of stars across the focal plane, the proposed algorithm estimates the angular rates in both camera frames, which can be translated into spacecraft inertial rates provided an attitude reference. The method is validated through numerical simulation and its accuracy assessed for a synthetic dataset of pointing and rates conditions, demonstrating its potential to complement conventional rate sensors for spacecraft systems or serve as a backup.

This paper is structured as follows. Section \ref{sec:methodology} describes the methodology for event-based angular rate estimation. Section \ref{sec:simulation} outlines the simulation approach and its implementation. Section \ref{sec:performance} evaluates the estimation accuracy using a dataset of synthetic event data. Finally, Section \ref{sec:conclusions} summarizes the main findings and conclusions of this work. 

\section{Methodology} \label{sec:methodology}
This section describes the methodology for estimating spacecraft angular rates from event-based data generated by the apparent motion of stars. The approach includes the definition of the relevant reference frames, the formulation of the attitude kinematics, the modeling of the event-based camera and event streams, and the procedures for motion field estimation and angular rate determination.
\subsection{Reference Frames}
Let us consider the International Celestial Reference Frame (ICRF). The origin of this reference frame lies in the Solar System Barycentre (SSB) and its axes are fixed by the position of distant quasars and radio sources in the universe \citep{ma1998international}. These sources have negligible proper motion with respect to the SSB, and, therefore, render the ICRF an inertial frame along typical timescales of space missions. Let us further consider the J2000 reference frame, which is the realization of the ICRF at the epoch 2000-01-01 12:00:00 in terrestrial time. This inertial frame is denoted as $\bm{E}$ = [$\hat{\bm{e}}_1$, $\hat{\bm{e}}_2$, $\hat{\bm{e}}_3$], where $\hat{\bm{e}}_i$ with $i$ = 1,2,3 are the unitary directions of the J2000 frame. These directions are defined such that $\hat{\bm{e}}_1$ points towards the vernal equinox, $\hat{\bm{e}}_3$ towards the north celestial pole, and $\hat{\bm{e}}_2$ completes the right-handed triad. These unitary directions are fixed in typical timescales of space missions. The unitary directions to stars in the universe can be assumed to be fixed with respect to this frame for typical mission durations, owing to the negligible relative angular motion along short time scales \citep{lefferts1982kalman}. The line-of-sight direction to stars in the inertial frame can be described according to their right ascension $\alpha$ and declination $\delta$ over the celestial sphere, which have been estimated by ground-based and space-based surveys \citep{kaiser2002pan}. These have been catalogued such as in the HIPPARCOS \citep{perryman1997hipparcos} or the GAIA catalogues \citep{vallenari2023gaia}. It can be noted that, while the 2-parameters model with $\alpha$ and $\delta$ will be used in the context of this paper, the accurate 5-parameters model including the stars proper motion can be used for precise astrometry studies \citep{christian2019starnav}. 

Let us now consider a spacecraft body frame denoted as $\bm{B}$ = [$\hat{\bm{b}}_1$, $\hat{\bm{b}}_2$, $\hat{\bm{b}}_3$], where $\hat{\bm{b}}_i$ with $i$ = 1,2,3 are the unitary directions along the principal axes of inertia of the spacecraft. The origin of this frame is located at the center of mass of the spacecraft. Since the parallax of stars is negligible for spacecraft located inside the solar system, we can assume that the line-of-sight directions to stars are not affected by the spacecraft position within the solar system. Therefore, the unitary direction to a star can be expressed in both frames as
\begin{equation} \label{eq:frames}
\bm{E} \, {\bm{{s}}}^{E} = \bm{B} \, {\bm{{s}}}^{B}
\end{equation}
where ${\bm{{s}}}^{E}$ and ${\bm{{s}}}^{B}$ are the normalized coordinates of a star direction in the inertial and body frames, respectively. From Eq.~\eqref{eq:frames}, the change in coordinates between reference frames can be easily obtained as
\begin{equation} \label{eq:frames2}
{\bm{{s}}}^{E} = \bm{R}_{EB} \, {\bm{{s}}}^{B} \qquad ; \qquad {\bm{{s}}}^{B} = \bm{R}_{BE} \, {\bm{{s}}}^{E}
\end{equation}
with $\bm{R}_{EB}$ = $\bm{E}^\top \bm{B}$ and $\bm{R}_{BE}$ = $\bm{B}^\top \bm{E}$ being the rotation matrices from the body to the inertial frames, and from the inertial to the body frames, respectively. These rotations, which are expressed through direction cosine matrices, usually leverage the definition of the Euler angles $\psi$, $\theta$, and $\phi$, denoting the twelve possible combinations of yaw, pitch, and roll angle rotations between frames \citep{shuster2006quest}. 

{We can now introduce a camera reference frame $\bm{C}$. In principle, the camera reference frame is not coincident with the body frame. In such a case, the coordinate transformations follow the equation
\begin{equation} \label{eq:frames3}
{\bm{{s}}}^{C} = \bm{R}_{CB} \, {\bm{{s}}}^{B} \qquad ; \qquad {\bm{{s}}}^{B} = \bm{R}_{BC} \, {\bm{{s}}}^{C}
\end{equation}
with $\bm{R}_{CB}$ = $\bm{C}^\top \bm{B}$ and $\bm{R}_{BC}$ = $\bm{B}^\top \bm{C}$ being the rotation matrices from the body to the camera frames, and from the camera to the body frames, respectively. The overall transformation from the inertial frame to the camera frame is thus given by ${\bm{{s}}}^{C} = \bm{R}_{CB} \, \bm{R}_{BE} \, {\bm{{s}}}^{E}$. Note that the orientation of the camera frame with respect to the body frame is typically known by the spacecraft configuration.} 

For simplicity, however, let us assume that $\bm{C}$ is coincident with the body frame $\bm{B}$ and shares the same origin. The normalized coordinates of a star in this frame are $\bm{{s}}^C$ = [$X$, $Y$, $Z$]$^\top$ and the corresponding line-of-sight direction in the camera frame can be expressed as $\bm{C} \, \bm{{s}}^C$. Let us now consider an imaging sensor installed in $\bm{C}$ which is modeled as a pin-hole camera \citep{ma2004invitation}. In this model, the coordinates of a star in $\bm{C}$ are projected on the focal plane of the sensor through perspective geometry as
\begin{equation} \label{eq:pinhole}
\left\{
\begin{array}{rl}
x &= f \, X/Z\\
y &= f \, Y/Z\\
z &= f
\end{array}
\right.
\end{equation}
where the point $\bm{p}^C$ = ($x$, $y$, $z$) is the projection of the star on the focal plane and $f$ is the camera focal length. Note that, for the definitions used in this paper, $||\bm{C} \, \bm{{s}}^C||$ = 1 and the distance of $\bm{p}^C$ to the origin is $f/Z$. 

\begin{figure}[htbp]
    \centering
    \includegraphics[width=0.95\linewidth]{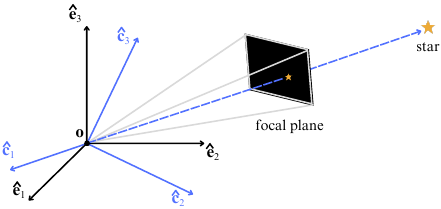}
    \caption{Visualization of the inertial reference frame [$\hat{\bm{e}}_1$, $\hat{\bm{e}}_2$, $\hat{\bm{e}}_3$], camera reference frame [$\hat{\bm{c}}_1$, $\hat{\bm{c}}_2$, $\hat{\bm{c}}_3$], and projection of a star onto the focal plane of a pinhole camera. For simplicity, the inertial and camera reference frames share the same origin owing to the negligible parallax of stars.}
    \label{fig:ref_frames}
\end{figure}

\subsection{Attitude Kinematics} \label{sec:attitude_kinematics}

Let us consider the case of a spacecraft in rotational motion about a given axis and at a given instantaneous rate. Following Eq.~\eqref{eq:frames}, the line-of-sight directions to the stars in the inertial frame and in the camera frame are related as
\begin{equation} \label{eq:framesfixedcamera}
\bm{E} \, {\bm{{s}}}^{E} = \bm{C} \, {\bm{{s}}}^{C}
\end{equation}
The apparent instantaneous rate of change of the line-of-sight directions to the stars can be obtained considering the time derivative of Eq.~\eqref{eq:framesfixedcamera} as
\begin{equation} \label{eq:imgrates}
\bm{0} = \dot{\bm{C}} \bm{{s}}^C + \bm{C} \dot{\bm{{s}}}^C \quad \rightarrow \quad \dot{\bm{{s}}}^C = - \bm{C}^\top \dot{\bm{C}} \, \bm{{s}}^C 
\end{equation}
where $\dot{\bm{{s}}}^C = [V_X, V_Y, V_Z]^\top$ is the apparent rate of change of the star coordinates in the camera frame and $\bm{\Omega} = \bm{C}^\top \dot{\bm{C}}$ is the skew-symmetric matrix corresponding to the angular velocity vector expressed in the camera frame, for which
\begin{equation} \label{eq:angularrates}
\bm{\Omega} = 
\begin{bmatrix}
0 & -r & q \\
r & 0 & -p \\
-q & p & 0
\end{bmatrix}
\end{equation}
where p, q, and r are the angular rates expressed in the camera frame. Therefore, considering Eq.~\eqref{eq:angularrates}, we can expand Eq.~\eqref{eq:imgrates} as
\begin{equation} \label{eq:imgrates2}
\left\{
\begin{array}{rl}
V_X &= + rY - qZ\\
V_Y &= - rX + pZ \\
V_Z &= +qX - pY
\end{array}
\right.
\end{equation}
Eq.~\eqref{eq:imgrates2} represents the apparent motion of a star in the camera frame due to spacecraft angular rotation. The rate of change of the star coordinates on the focal plane, denoted by the apparent velocities u, v, and w, can be derived by differentiating Eq.~\eqref{eq:pinhole} as
\begin{equation} \label{eq:pinhole2}
\left\{
\begin{array}{rl}
u &= f \, (V_X Z - V_Z X) / Z^2\\
v &= f \, (V_Y Z - V_Z Y) / Z^2 \\
w &= 0
\end{array}
\right.
\end{equation}
We can further develop Eq.~\eqref{eq:pinhole2} considering the pinhole transformations in Eq.~\eqref{eq:pinhole} and the apparent motion in Eq.~\eqref{eq:imgrates2}, resulting in
\begin{equation}
\begin{aligned}
u &= \frac{1}{f} \left[ r y f + p x y - q (f^2 + x^2) \right] \\
v &= \frac{1}{f} \left[ -r x f + p (f^2 + y^2) - q x y \right]
\end{aligned}
\end{equation}
with $w$ being unobservable as it is always null, coming from the pinhole model equation ($z = f$). These expressions describe how the image coordinates of a fixed star move on the image plane as the camera rotates with angular rates (p, q, r). Note that these expressions can also be written in a more compact matrix form as
\begin{equation} \label{eq:motionfield}
\begin{bmatrix}
u \\
v
\end{bmatrix}
=
\frac{1}{f}
\begin{bmatrix}
 x y & -(f^2 + x^2) & y f \\
f^2 + y^2 & -x y & -x f
\end{bmatrix}
\begin{bmatrix}
p \\
q \\
r
\end{bmatrix}
\end{equation}
where we can identify the matrix
\begin{equation}
\bm{F}(x, y) = \frac{1}{f}
\begin{bmatrix}
x y & -(f^2 + x^2) & y f \\
f^2 + y^2 & -x y & -x f
\end{bmatrix}
\end{equation}
commonly known in literature as feature sensitivity matrix \citep{shiba2023fast}, such that 
\begin{equation} \label{eq:motionfield2}
\begin{bmatrix}
u \\
v
\end{bmatrix}
=
\bm{F}(x, y)
\begin{bmatrix}
p \\
q \\
r
\end{bmatrix}
\end{equation}
Eq.~\eqref{eq:motionfield2} maps the spacecraft angular rates in the camera frame into the apparent motion field of stars.
\subsection{Angular Rate Estimation}
Eq.~\eqref{eq:motionfield2} can be written for each pixel. Therefore, given a set of N measurements ($x_i$, $y_i$, $u_i$, $v_i$) for i = 1, .., N, the angular rates (p, q, r) can be estimated in a least squares sense solving the system:
\begin{equation} \label{eq:leastsquares}
\underbrace{\begin{bmatrix}
u_1 \\
v_1 \\
\vdots \\
u_i \\
v_i \\
\vdots \\
u_\textrm{N} \\
v_\textrm{N}
\end{bmatrix}}_{\bm{y}}
=
\underbrace{\begin{bmatrix}
\bm{F}(x_1, y_1) \\
\vdots \\
\bm{F}(x_i, y_i) \\
\vdots \\
\bm{F}(x_\textrm{N}, y_\textrm{N})
\end{bmatrix}}_{\bm{H}}
\underbrace{\begin{bmatrix}
p \\
q \\
r
\end{bmatrix}}_{\bm{x}}
\end{equation}
whose solution is $\bm{x} = (\bm{H}^\top \bm{H})^{-1}\bm{H}^\top \, \bm{y}$. {Note that, in order to solve Eq.~\eqref{eq:leastsquares}, the matrix $\bm{H}$ must be full rank. The problem is underdetermined in the case of a single star, and $\bm{H}$ may become ill-conditioned or rank-deficient when the observed stars exhibit degenerate spatial configurations (e.g., when stars are poorly separated). This issue is mitigated through an appropriate selection of the camera field-of-view (FoV), based on the statistical number and spatial distribution of stars within the FoV, which ensures the presence of multiple well-separated stars at any given time. This consideration is analogous to the one for star trackers, which face the same observability constraints and therefore employ fields of view on the order of 10 degrees to guarantee robust attitude and angular rate estimation.} Now, considering $\bm{y}$ to be affected by a measurement error $\Delta \bm{y}$ and the solution error to be $\Delta \bm{x}$, it is easy to verify that the analytical covariance reads:
\begin{equation}
\bm{R}_{xx} = (\bm{H}^\top\bm{H})^{-1}\bm{H}^\top \bm{R}_{yy} \, \bm{H} \, (\bm{H}^\top\bm{H})^{-\top}
\end{equation}
where $\bm{R}_{xx} = \textrm{E}[\Delta \bm{x} \Delta \bm{x}^\top]$ denotes the solution covariance, $\bm{R}_{yy} = \textrm{E}[\Delta \bm{y} \Delta \bm{y}^\top]$ is the measurement covariance, and $\textrm{E}$ denotes the expectation operator. Therefore, the problem is now to retrieve the set of measurements ($x_i$, $y_i$, $u_i$, $v_i$), them being the apparent motion field distribution over the pixel array of the event camera sensor.

\subsection{Event-Based Sensing}
Unlike traditional frame-based cameras that capture full images at fixed time intervals, event-based sensors detect changes in brightness asynchronously and independently for each pixel, resulting in a stream of events rather than a sequence of full images~\citep{delbruck2010activity}. An event is triggered at a single pixel when a logarithmic change in brightness exceeds a predefined threshold~\citep{gallego2020event}. An event is therefore defined as e = (x, y, t, k), where t denotes the timestamp, x and y mark the pixel location, and k, attaining values $\pm$ 1, indicates whether the pixel intensity increased or decreased according to a logarithmic difference threshold, i.e, 
\begin{equation} \label{eq:intensity}
\begin{aligned}
\textrm{log} \, \textrm{I} \, (x, y, t) - \textrm{log} \, \textrm{I} \, (x, y, t^*) & \geq + e_{\textrm{t}} \quad \rightarrow \quad k = + 1 \\
\textrm{log} \, \textrm{I} \, (x, y, t) - \textrm{log} \, \textrm{I} \, (x, y, t^*) & \leq -e_{\textrm{t}} \quad \rightarrow \quad k = - 1
\end{aligned}
\end{equation}
where $\textrm{I}$ indicates the pixel brightness, $t^*$ denotes the last time the event was triggered on the pixel (x, y), and $e_{\textrm{t}}$ corresponds to the event threshold according to the camera specifications. In the context of this paper, event-based sensing is employed to retrieve the apparent motion field ($u_i$, $v_i$) of the stars.

\subsection{Event Stream Representation}
\begin{figure}[htbp]
  \centering
  \begin{subfigure}[b]{0.4\linewidth}
    \centering
    \includegraphics[width=\textwidth]{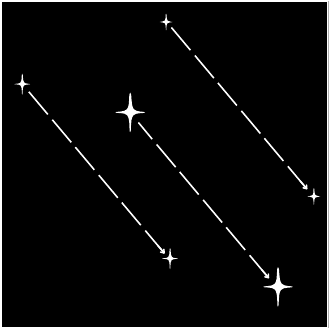}
    \caption{}
  \end{subfigure}
  \hspace{0.3cm}
  \begin{subfigure}[b]{0.4\linewidth}
    \centering
    \includegraphics[width=\textwidth]{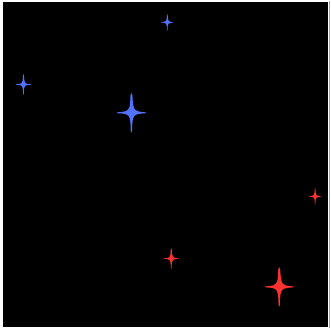}
    \caption{}
  \end{subfigure}
  \caption{Apparent motion of stars on the focal plane of (a) a frame-based camera and (b) an event-based camera. The red and blue pixels denote positive and negative event polarity.}
  \label{fig:starpolarity}
\end{figure}

\begin{figure}[htbp]
  \centering
  \begin{subfigure}[b]{0.49\linewidth}
    \centering
    \includegraphics[width=\textwidth]{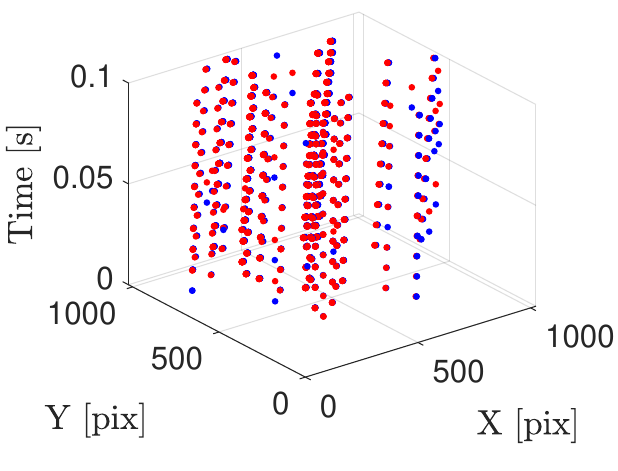}
    \caption{} \label{fig:eventsrepresentationa}
  \end{subfigure}
  \hspace{0.01cm}
  \begin{subfigure}[b]{0.48\linewidth}
    \centering
    \includegraphics[width=\textwidth]{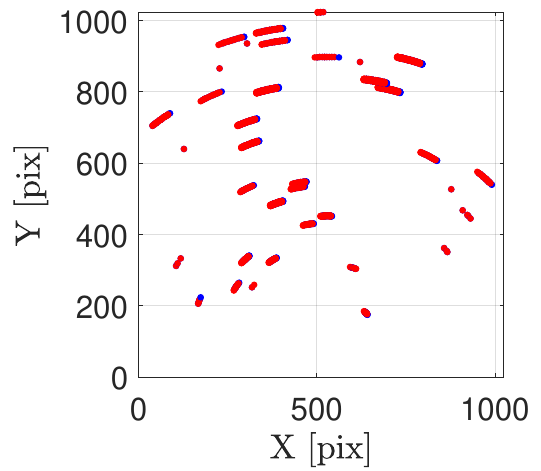}
    \caption{} \label{fig:eventsrepresentationb}
  \end{subfigure}  \\
  \begin{subfigure}[b]{0.48\linewidth}
    \centering
    \includegraphics[width=\textwidth]{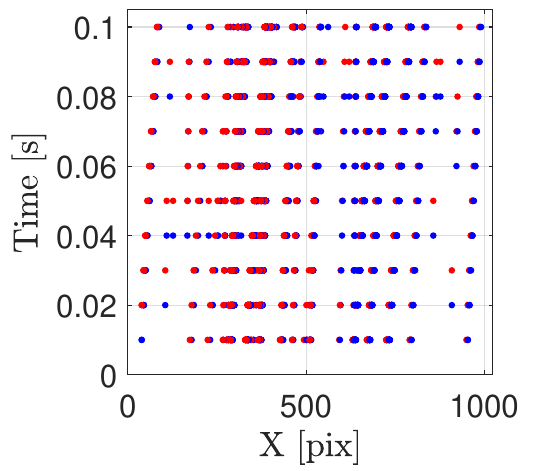}
    \caption{} \label{fig:eventsrepresentationc}
  \end{subfigure} 
  \hspace{0.01cm}
  \begin{subfigure}[b]{0.48\linewidth}
    \centering
    \includegraphics[width=\textwidth]{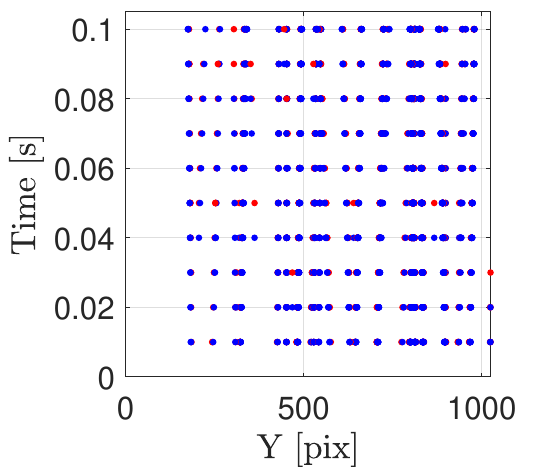}
    \caption{} \label{fig:eventsrepresentationd}
  \end{subfigure} 
  \caption{Representation of a stream of events: (a) events triggered by the apparent motion of stars, which are shown as dotted lines in the (x, y, t) space; (b) projection of events on the (x, y) plane; (c) projection of events on the (x, t) plane; (d) projection of events on the (y, t) plane. The projections of events on the (x, y) plane reveal the apparent motion of stars, while the projections on the (x, t) and (y, t) planes highlights the apparent motion field components u and v, respectively.}
  \label{fig:eventsrepresentation}
\end{figure}

Let us consider the case of a spacecraft under instantaneous rotational motion about a given axis and at a given rate as in Section \ref{sec:attitude_kinematics}. The spacecraft is equipped with an event-based sensor and, for simplicity, let us assume that the body frame matches the camera frame. While rotating, the camera continuously provides event locations according to the threshold in Eq.~\eqref{eq:intensity} in the form $e_i$ = ($x_i$, $y_i$, $t_i$, $k_i$). Stars, which appear to move across the field of view, trigger events along their paths, as shown in Figure \ref{fig:starpolarity}. 
Let us consider a local region of pixels surrounding a star. As the star moves across the field of view, this region is activated by events of positive polarity along the apparent direction of motion, and by events of negative polarity in the opposite direction. The coexistence of opposite-polarity events within the same local region encodes information about the direction of motion, while the separation between these polarities is proportional to the magnitude of the motion field. Fig.~\ref{fig:eventsrepresentation} illustrates a sample event stream triggered by a rotating camera. The stream is visualized in the spatiotemporal (x, y, t) space (Fig.~\ref{fig:eventsrepresentationa}) and projected onto three orthogonal planes. The projection onto the (x, y) plane highlights the apparent motion of the stars over time (Fig.~\ref{fig:eventsrepresentationb}), while the projections onto the (x, t) and (y, t) planes reveal the components of the apparent velocity, u and v, represented as the slopes of the event trajectories (Fig.~\ref{fig:eventsrepresentationc} and Fig.~\ref{fig:eventsrepresentationd}, respectively). 

{Note that, without loss of generality, this paper considers event acquisitions lasting up to 0.1 s. It is important to note that event cameras can continuously acquire for longer time periods with a time resolution down to the microsecond. Typical maximum rates for agile (or tumbling) spacecraft can reach values of 20-30 deg/s, while controlled rates for common operations are in the order of 1-2 deg/s or lower \citep{Markley2014spacecraft}. This has led to a measurement window of 0.1 s while assuming a constant spacecraft angular rate during this window, corresponding to angular rotations of 2-3 deg and 0.1-0.2 deg, respectively, in the mentioned cases. This window ensures that enough events are triggered given the considered rates, leading to the applicability of the method to common slew maneuvers. A shorter maximum time window can be selected for faster operations, and a longer window can be selected for slower maneuvers. Note also that an analyst is free to devise an adaptive duration window according to the event stream detected.} 

\subsection{Motion Field Estimation}
\begin{figure}[htbp]
  \centering
  \begin{subfigure}[b]{0.47\linewidth}
    \centering
    \includegraphics[width=\textwidth]{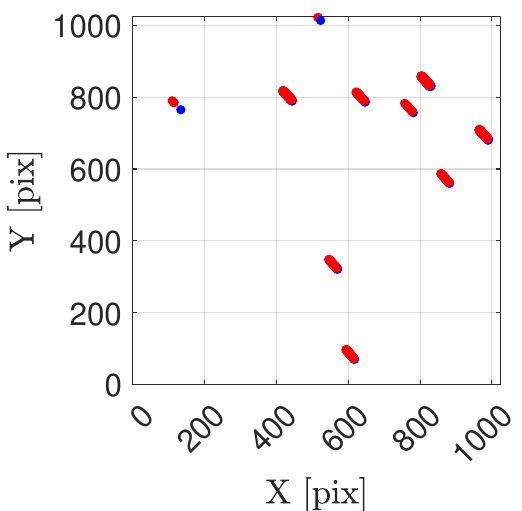}
    \caption{} \label{fig:unwarped}
  \end{subfigure}
  \hspace{0.1cm}
  \begin{subfigure}[b]{0.47\linewidth}
    \centering
    \includegraphics[width=\textwidth]{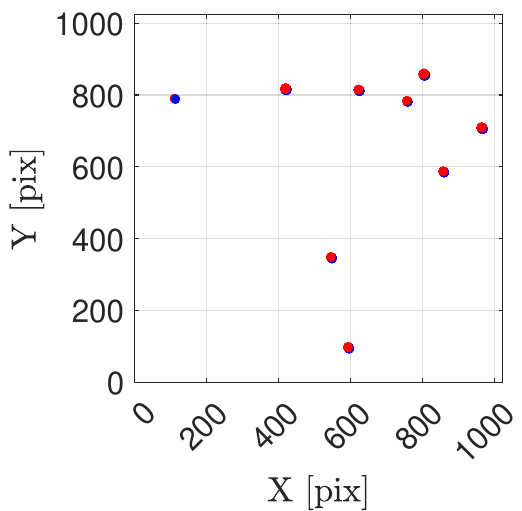}
    \caption{} \label{fig:warped}
  \end{subfigure}
  \caption{Contrast maximization algorithm: (a) Input unwarped events stream represented on the (x, y) plane; (b) Output warped event stream on the (x, y) plane according to the best fit velocity $\bm{v}^*$ = [$u$, $v$].}
  \label{fig:contrastmaximization}
\end{figure}

Figure \ref{fig:unwarped} shows typical events triggered by the apparent motion of the stars on the focal plane of the camera. If we could warp back in time all the events with the correct motion field of stars, we would obtain a sharp image as the one shown in Figure \ref{fig:warped}. This is the fundamental principle of the contrast maximization algorithm \citep{guo2024cmax}. Given an event stream ($x_i$, $y_i$, $t_i$, $k_i$), the contrast maximization algorithm estimates a global motion field vector $\bm{v} = [u, v]$ to maximize the sharpness of the image formed by the accumulated events. This is achieved by folding back to a reference time all the events according to the estimated $\bm{v}$. The optimal $\bm{v}^*$ is the one that maximizes the sharpness of the event-accumulated image.

Let us denote by \( \mathcal{V} = \{(x_i, y_i, t_i, k_i)\}_{i=1}^N \) the sequence of \( N \) events generated by a rotating camera, and by \( \textrm{H} \) and \( \textrm{W} \) the image height and width from the camera model. Let the reference time be defined as \( t_0 = \min t_i \). The contrast maximization algorithm follows a two-stage approach: an initialization phase and a fine optimization step. A coarse grid-search evaluates a set of candidate image velocities  \( \bm{v}_g = [u_g, v_g] \) within a predefined range. For each candidate \( \bm{v}_g \), the contrast score over the grid is computed, and the best-performing velocity \( \bm{v}_{\text{c}} \) is retained. The contrast score, denoted \( S(u, v) \), is computed by warping events to the reference time \( t_0 \) based on the motion model:
\begin{equation}
    x_i^w = x_i - u  \, (t_i - t_0), \quad
    y_i^w = y_i - v \, (t_i - t_0)
\end{equation}
The warped events \( (x_i^w, y_i^w) \) are then accumulated into an image \( I \) by discretizing into pixels and summing the polarities. Therefore, each accumulated pixel will have a value:
\begin{equation}
    I_{i}(u,v) = \sum_{\substack{(x_i^w, y_i^w)}} k_i
\end{equation}
The contrast score \( S(u, v) \) is defined as the variance of the image \( I(u,v) \) with mean $\mu$:
\begin{equation}
    S(u, v) = \frac{1}{\Omega} \sum_{\substack{(x_i^w, y_i^w)}} (I_i(u,v) - \mu) ^2
\end{equation}
where $\Omega$ is the overall number of pixels. A higher contrast score \( S(u, v) \) indicates better alignment of events under the assumed candidate velocity, resulting in a sharper accumulated image. After the grid search, the best-performing velocity \( \bm{v}_{\text{c}} \) is used as the starting point for a refined optimization step. We define the objective function as:
\begin{equation}
    J(u, v) = -S(u, v)
\end{equation}
Note that different contrast/loss functions exist \citep{gallego2019focus}. The goal is to minimize \( J(u, v) \), which corresponds to maximizing the contrast. In this work, a Newton method is used to find the local minimum. After convergence, the optimal velocity \( \bm{v}^* = [u^*, v^*] \) is obtained and used to solve Eq.~\eqref{eq:leastsquares}. The algorithm assumes a constant image velocity field over the duration of the event interval, which is a valid assumption for the short time frames considered. {Note that, in principle, a per-pixel set of images velocities $(u_i, v_i)$ is required to solve Eq.~\eqref{eq:leastsquares}. However, this requires tracking, clustering, and applying contrast maximization to local patches associated to tens of stars, slowing down the process. In our case, the global velocity $[u^*, v^*]$ obtained by contrast maximization is used as estimation of each per-pixel entry in Eq.~\eqref{eq:leastsquares}, knowing that a small per-pixel deviation is introduced. The equations still hold, as the $\bm{F}(x_i, y_i)$ matrix is still different per each pixel, leading to an estimation of (p, q, r) in a least squares sense. For this to work, the $\bm{H}$ matrix, which is constructed by stacking the $\bm{F}(x_i, y_i)$ matrices, has to be full rank, which is ensured by the spreadness and number of stars considered within the field of view. Note that, while estimating a per-pixel set of velocities $(u_i,v_i)$ would lead to higher accuracy, using a global $[u^*,v^*]$ still leads to accurate performance in virtue of a fast processing time.}

\section{Simulation} \label{sec:simulation}

\begin{figure}[htbp]
    \centering
    \includegraphics[width=0.9\linewidth]{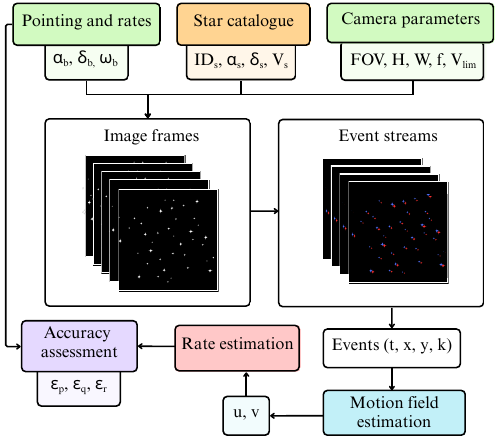}
    \caption{Block diagram of the simulation workflow for event-based angular rate estimation.}
    \label{fig:sim_block}
\end{figure}

The simulation workflow is illustrated in Figure~\ref{fig:sim_block}. The process is divided into six main functional blocks: 1) star catalogue retrieval, 2) spacecraft camera, pointing, and rates definition, 3) image frame generation, 4) events stream generation, 5) angular rate estimation, and 6) accuracy assessment.
The stellar data are retrieved from the HIPPARCOS and GAIA catalogues. The HIPPARCOS catalogue contains approximately 118,000 sources down to a visual magnitude of 12.4, while GAIA DR3 includes about 1.81 billion sources down to mag 20.7. These catalogues are parsed to extract the star identifiers (ID$_\textrm{s}$), right ascension ($\alpha_\textrm{s}$), declination ($\delta_\textrm{s}$), and magnitude (V$_\textrm{s}$). These parameters define an inertial sky map as observed from a spacecraft located within the solar system, under the assumption of negligible parallax and proper motion effects over the simulation timescale. The unitary star coordinates in the inertial frame are expressed as 
\begin{equation} \label{eq:sphericalcoordinates}
\bm{{s}}^E = [\textrm{cos}(\delta_\textrm{s}) \, \textrm{cos}(\alpha_\textrm{s}) \,,\, 
\textrm{cos}(\delta_\textrm{s}) \, \textrm{sin}(\alpha_\textrm{s})\,,\, \textrm{sin}(\delta_\textrm{s})]^\top
\end{equation}
Then, the spacecraft’s camera parameters and motion states are defined. This includes the camera's field of view, sensor array dimensions (H, W), focal length (f), and limit magnitude (V$_\textrm{lim}$). For each simulation case, a set of random boresight directions ($\alpha_\textrm{b}$, $\delta_\textrm{b}$) and angular rates ($\bm{\omega}_\textrm{b}$) is generated to simulate various spacecraft orientations and rotational motions. The boresight directions are defined evaluating Eq.~\eqref{eq:sphericalcoordinates} with $\alpha_\textrm{b}$ and $\delta_\textrm{b}$, while the rotational motion is defined according to Eq.~\eqref{eq:imgrates} and Eq.~\eqref{eq:imgrates2}. The boresight directions span the whole celestial sphere ($\alpha_\textrm{b} \in$ [0, 360] deg, $\delta_\textrm{b} \in$ [-90, 90] deg), while the angular motion is randomly set between -30 deg/s and 30 deg/s for each of the three rate components.  In this way, the star catalogues and spacecraft configuration parameters are used to synthetize a sequence of image frames containing stars. These frames simulate the apparent motion of stars across the camera sensor caused by the spacecraft's angular rotation. The star coordinates are projected on the frames according to the pin-hole model in Eq.~\eqref{eq:pinhole}. The event streams are then generated from the simulated image frames. {Note that there is no need to identify stars or star patterns in this pipeline, as in star trackers, since the observables here are the event streams generated by a consistent angular motion of the spacecraft.} Events are detected by identifying per-pixel logarithmic changes in brightness over time according to Eq.~\eqref{eq:intensity}, emulating the behaviour of event-based vision sensors. This produces asynchronous streams of events ($x$, $y$, $t$, $k$) for each boresight and angular rate scenario. The angular rate estimation algorithm, described in Section~\ref{sec:methodology}, can now be implemented, which uses the event data to infer the motion field and in turn estimate the spacecraft’s angular velocity solving Eq.~\eqref{eq:leastsquares}. Finally, the estimated angular rates are compared to the ground truth values defined in the configuration, allowing for the quantification of estimation accuracy across the simulation cases. The instantaneous angular rate errors are defined as $\varepsilon_p$, $\varepsilon_q$, and $\varepsilon_r$, which are used to synthesize the accuracy metric $\varepsilon_{\textrm{RMS}}$ as
\begin{equation}
\varepsilon_{\textrm{RMS}} = \frac{1}{\textrm{N}} \sqrt{\sum_1^\textrm{N} (\varepsilon_p^2 + \varepsilon_{{q}}^2 + \varepsilon_r^2)}
\end{equation}
it being the root mean square error of the angular rate estimation, which is evaluated per each simulation case, where $N$ is the number of evaluation points per each simulation. Note that the angular rates can be transformed into the inertial frame leveraging knowledge of the spacecraft attitude by other sensors such as star trackers.

\section{Performance} \label{sec:performance}

\begin{figure}[htbp]
    \centering
    \includegraphics[width=0.95\linewidth]{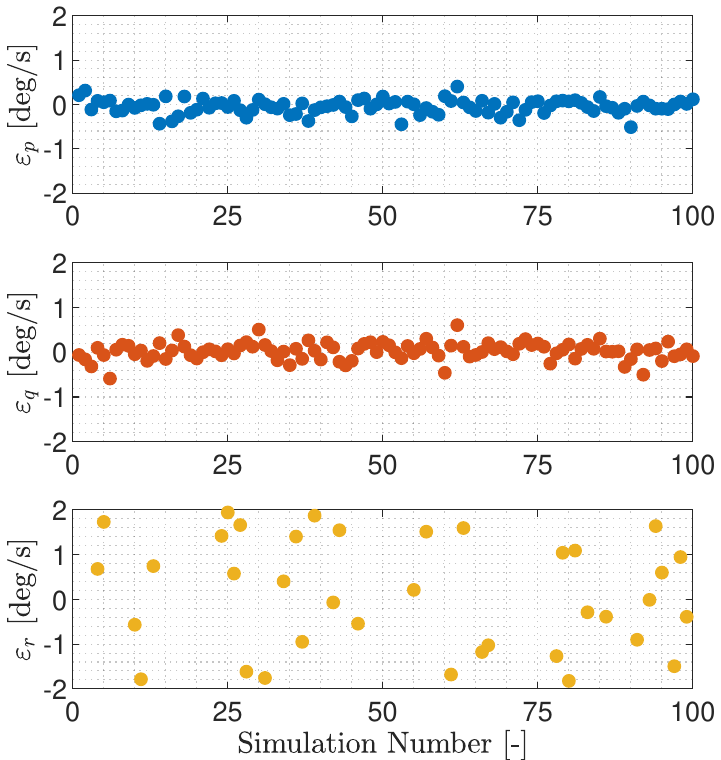}
    \caption{Estimation accuracy for $p$, $q$, and $r$ components. A sample set of 100 simulations is shown.}
    \label{fig:error_pqr}
\end{figure}

Figure~\ref{fig:error_pqr} shows the estimation accuracy of the angular velocity components $p$, $q$, and $r$ over a representative subset of 100 simulations, based on the methodology introduced in Section~\ref{sec:methodology} and Section~\ref{sec:simulation}. A random pointing ($\alpha_b$, $\delta_b$) and angular rate ($\bm{\omega}_b$) is set for each simulation and the corresponding rate estimates in the camera frame are plotted. We can note high accuracy in estimating the yaw ($q$) and pitch ($p$) rates, while the roll rate ($r$) is significantly less accurate. This behavior is due to the third velocity component in the image plane, $w$, which is not directly observable from 3D to 2D projections, as shown in Eq.~\eqref{eq:pinhole2}.

To improve the estimation performance along all three rotational axes, we consider a dual event camera setup with orthogonal mounting. This strategy is commonly adopted in miniaturized platforms as the use of orthogonal star trackers in CubeSat-sized spacecraft~\citep{franzese2022DeepSpace}. Deploying multiple sensors enhances the estimation accuracy, but also improves reliability, robustness, and redundancy of the estimation. Thanks to their low size, mass, and power (SWaP) requirements, event cameras are particularly suitable for this approach. For example, they can weigh as little as 40\,g, occupy a volume as small as $30\times30\times36$\,mm$^3$, and consume as little as 1\,W as peak power~\citep{chakravarthi2025recent}. Let us denote the reference frames of the two cameras as $\bm{C}_A = [{\bm{\hat c}_{A1}}, {\bm{\hat c}_{A2}}, {\bm{\hat c}_{A3}}]$ and $\bm{C}_B = [{\bm{\hat c}_{B1}}, {\bm{\hat c}_{B2}}, {\bm{\hat c}_{B3}}]$. Without loss of generality, we choose an orthogonal configuration such that:
\begin{equation} \label{eq:ortmounting}
\left\{
\begin{array}{rl}
{\bm{\hat c}_{B1}} &= {\bm{\hat c}_{A1}} \\
{\bm{\hat c}_{B2}} &= - {\bm{\hat c}_{A3}} \\
{\bm{\hat c}_{B3}} &= {\bm{\hat c}_{A2}} \\
\end{array}
\right.
\end{equation}
In this configuration, the roll axis of the first camera (${\bm{\hat c}_{A3}}$) is aligned (but opposite) to the yaw axis of the second camera (${\bm{\hat c}_{B2}}$), and the roll axis of the second camera (${\bm{\hat c}_{B3}}$) is aligned with the yaw axis of the first camera (${\bm{\hat c}_{A2}}$). This setup allows us to reject roll rate measurements (which are poorly estimated) from both cameras, and instead fuse pitch and yaw rate information to obtain accurate angular velocity estimation along all three axes. An example of fusion strategy to estimate the angular velocity in the first camera frame is:
\begin{equation} \label{eq:ortmounting3}
\left\{
\begin{array}{rl}
p_f &= \frac{1}{2} \, (p_1 + p_2)\\
q_f &= q_1 \\
r_f &= -q_2 \\
\end{array}
\right.
\end{equation}
where $(p_f, q_f, r_f)$ are the fused angular rate estimates in the $\bm{C}_A$ frame, and $(p_i, q_i, r_i)$ for $i=1,2$ are the measurements from the two event cameras. Note that equal weighting between $p_1$ and $p_2$ is adopted here since both cameras are assumed to be identical; more general weighted fusion strategies can be implemented when selecting different cameras. Also, the fused rates can be expressed in the second camera frame:
\begin{equation} \label{eq:ortmounting4}
\left\{
\begin{array}{rl}
p_f &= \frac{1}{2} \, (p_1 + p_2) \\
q_f &= q_2 \\
r_f &= q_1 \\
\end{array}
\right.
\end{equation}

\begin{figure}[htb]
    \centering
    \includegraphics[width=0.98\linewidth]{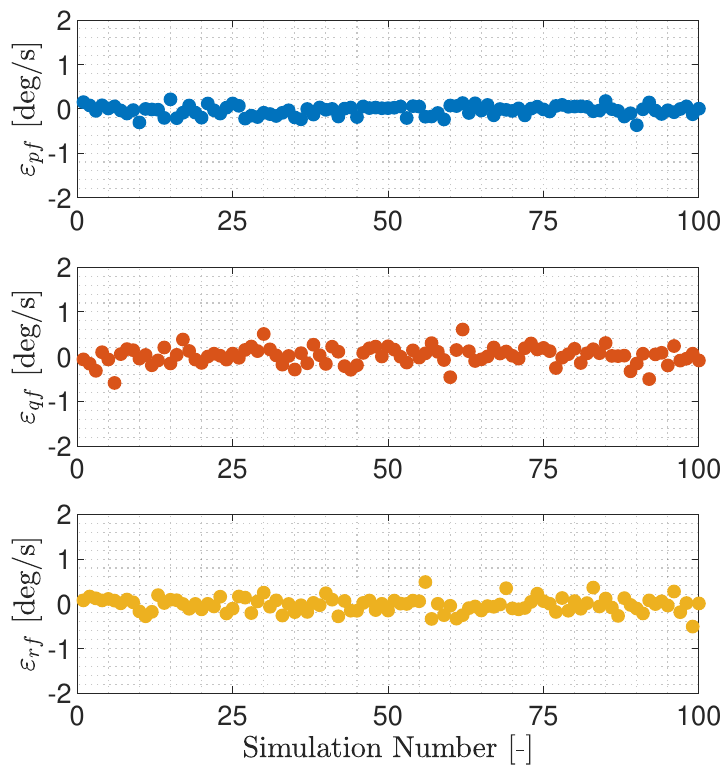}
    \caption{Estimation accuracy for $p$, $q$, and $r$ components after sensor fusion. A sample set of 100 simulations is shown.}
    \label{fig:error_f}
\end{figure}

Figure~\ref{fig:error_f} shows the estimation accuracy for all three angular rates after sensor fusion using the dual-camera configuration. The estimation is significantly improved, achieving an accuracy along the roll axis similar to the one along the pitch and yaw axes.

Table~\ref{tab:rms_error} summarizes the root mean square (RMS) errors for single and dual camera configurations. With the orthogonal dual-camera setup, a total RMS error of 0.0275\,deg/s is achieved. {This performance slightly outperforms the typical accuracy of miniaturized inertial measurement units (0.05–0.5 deg/s), which are also affected by bias instability and drift and are widely used on small satellites. It is also competitive with angular rate estimates derived from frame-based star trackers (0.01–0.1 deg/s) \citep{wertz2012spacecraft}. The event-based approach offers advantages during fast slew rate maneuvers, where frame cameras fail due to motion blur, and provides high rate accuracy in high dynamic range environments near bright objects (e.g., in Earth proximity). The event camera is also a low-latency and low-power sensor that can be used for other tasks in addition to rate estimation.} 

\begin{table*}[htb!]
\centering
\begin{tabular}{lcccc}
\toprule
\textbf{Configuration} & $\boldsymbol{\varepsilon_p}$ [deg/s] & $\boldsymbol{\varepsilon_q}$ [deg/s] & $\boldsymbol{\varepsilon_r}$ [deg/s] & \textbf{Total RMS} [deg/s] \\
\midrule
Single Camera  & 0.0165 & 0.0192 & 0.3060 & 0.3070 \\
Double Camera  & 0.0115 & 0.0192 & 0.0160 & 0.0275 \\
\midrule
\textbf{Reference Frame} & $\boldsymbol{\varepsilon_{\omega x}}$ [deg/s] & $\boldsymbol{\varepsilon_{\omega y}}$ [deg/s] & $\boldsymbol{\varepsilon_{\omega z}}$ [deg/s] & \textbf{Total RMS} [deg/s] \\
\midrule
Inertial       & 0.0148 & 0.0143 & 0.0183 &  0.0275    \\
\bottomrule
\end{tabular}
\caption{Root mean square (RMS) error for angular rate estimation in single and dual camera configurations.}
\label{tab:rms_error}
\end{table*}

\begin{figure}[htbp]
    \centering
    \includegraphics[width=0.98\linewidth]{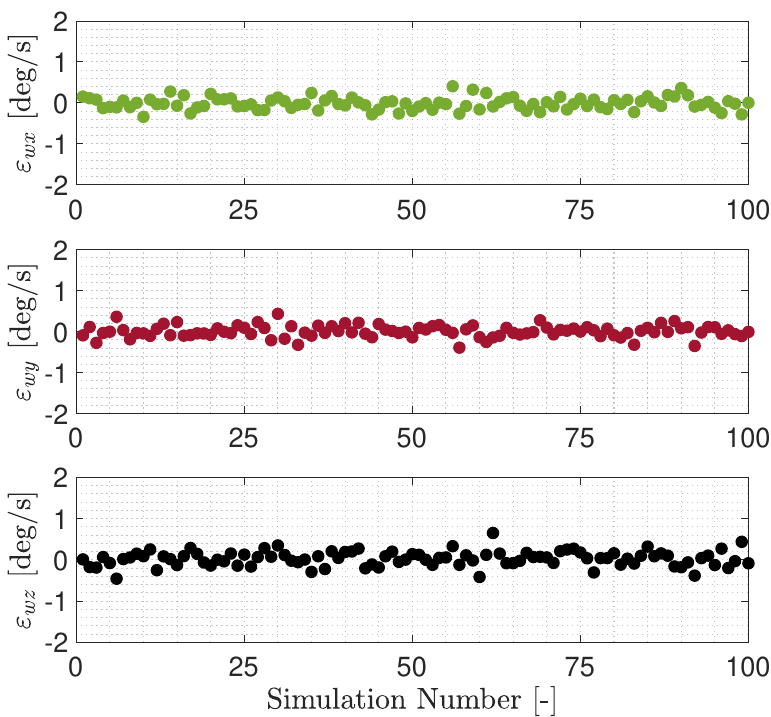}
    \caption{Estimation accuracy for $\omega_x$, $\omega_y$, and $\omega_z$ components in the inertial frame after sensor fusion. A sample set of 100 simulations is shown.}
    \label{fig:error_w}
\end{figure}

At this stage, we can also leverage knowledge of the spacecraft attitude by other sensors, such as star trackers, and the camera mounting configuration, to transform the fused angular velocity measurements into the inertial frame. The resulting estimation accuracy in the inertial angular velocity components, denoted as $\varepsilon_{wx}$, $\varepsilon_{wy}$, and $\varepsilon_{wz}$, is shown in Figure~\ref{fig:error_w}. As summarized in Table \ref{tab:rms_error}, the root mean square error per inertial rate components is better than 0.019 deg/s, with an overall root mean square error of 0.0275 deg/s across the simulations.

{It is now beneficial to evaluate the sensitivity of the estimation accuracy to noise models injected in the simulation. To this aim, we can consider four Gaussian noise sources in the estimation loop, respectively impacting the time stamps of the events, the pixel locations, the rate variation within the measurement window, and a residual misalignment between the two cameras after calibration. These sources are injected as Gaussian noise models in three cases of increasing uncertainty (C-1, C-2, and C-3), with the 3$\sigma$ standard deviation confidence bounds detailed in Table \ref{tab:uncertainties_models}. Pixel-level noise has been found to have the largest impact on the RMS accuracy among the sources considered, reaching a degradation of +4.95$\%$ for the largest uncertainty case ($\pm 0.5$ pixel). The timestamp uncertainty (up to $10 \mu$s, causing +3.07$\%$ degradation) and rate variation noise (up to $\pm 0.005$ deg/s, causing +2.23$\%$ degradation) produce marginal and acceptable increases in the RMS estimation accuracy. The residual inter-camera misalignment remains a limited effect within the post-calibration accuracies considered.}

\begin{table}[ht!]
\centering
\setlength{\tabcolsep}{4pt}
\begin{tabular}{lcccc}
\toprule
{{Noise source}} & {Unit} & {C-1} & {C-2} & {C-3} \\
\midrule
{Timestamp}   & {$\mu$s}    & {$\pm 1$} & {$\pm5$} & {$\pm 10$} \\
{Pixel location}  & {pix} & {$\pm 0.1$} & {$\pm 0.3$} & {$\pm 0.5$} \\
{Rate variation}   & {deg/s} & {$\pm 0.001$} & {$\pm 0.003$} & {$\pm 0.005$} \\
{Misalignment}     & {arcsec} & {$\pm 1$} & {$\pm 5$} & {$\pm 10$} \\
\bottomrule
\end{tabular}
\caption{{Noise sources and uncertainty levels (3$\sigma$) for the sensitivity analysis in three cases: C-1, C-2, and C-3.}}
\label{tab:uncertainties_models}
\end{table}

\begin{table}[ht!]
\centering
\begin{tabular}{lccc}
\toprule
{RMS Increase} & {C-1} & {C-2} & {C-3} \\
\midrule
{Timestamp}       & {0.12$\%$} & {1.21$\%$} & {3.07$\%$} \\
{Pixel location}  & {0.72$\%$} & {2.47$\%$} & {4.95$\%$} \\
{Rate variation}   & {0.36$\%$} & {1.08$\%$} & {2.23$\%$} \\
{Misalignment}     & {$0.01\%$} & {0.68$\%$} & {1.57$\%$} \\
\bottomrule
\end{tabular}
\caption{{Percent increase in RMS error due to injected noise models corresponding to the uncertainty levels in cases C-1, C-2, and C-3. Percentages are computed with respect to the nominal dual-camera RMS value of 0.0275 deg/s}}
\label{tab:uncertainties_effect}
\end{table}

\section{Conclusions} \label{sec:conclusions}

This paper presented a framework for determining spacecraft angular velocity using event-based camera sensing. This is achieved by leveraging the events triggered by the apparent motion of stars across the sensor focal plane. The method was validated through numerical simulations based on synthetic event streams generated from real star catalogues under randomized spacecraft boresight and angular rate configurations. Across a large dataset of test cases and relying on a two orthogonally mounted camera configuration, the approach achieved a root mean squared accuracy in angular rate estimation of 0.0275 deg/s along the three axes. {The sensitivity analysis with four noise sources has shown a marginal increase in the RMS accuracy bounded by 5$\%$ to the 0.0275 deg/s estimation}. These results highlight the potential of event-based vision for angular rate sensing in space, offering an alternative to, or complementing, conventional rate sensors such as gyroscopes. 

\section*{Acknowledgments} \label{sec:acknowledgements}

This research received no external funding. The authors would like to thank Dr.~Guillermo Gallego for inspiring discussions on event-based vision.

\bibliographystyle{elsarticle-harv} 
\bibliography{references}

\end{document}

%% file: Packages.tex
\usepackage{amssymb}
\usepackage{acro}
\usepackage{amsmath}
\usepackage{multirow}
\usepackage{comment}
\usepackage{tabularx}
\usepackage{multirow}
\usepackage{booktabs}
\usepackage{subcaption}
\usepackage{colortbl}
\usepackage{soul}
\usepackage{tikz}
\usepackage{textcomp}
\usepackage{longtable}
\usepackage{enumitem}
\usepackage{siunitx}
\usepackage{hyperref}
\usepackage{blindtext}
\usepackage{enumitem}
\definecolor{Gray}{gray}{0.9}

\usepackage{ulem}
\usepackage{mwe}

\usepackage{booktabs}
\usepackage{lineno}
\linenumbers

\journal{Advances in Space Research}

\usepackage{setspace}
\usepackage{algorithm}
\usepackage{algpseudocode}
\usepackage{algorithmicx}
\usepackage{mathtools}
\usepackage{threeparttable}
\usepackage[table]{xcolor}
\usepackage[version=4]{mhchem}
\usepackage{siunitx}
\usepackage{bm}

\makeatletter
\def\BState{\State\hskip-\ALG@thistlm}
\makeatother

\algnewcommand\algorithmicsystem{\textbf{System}}
\algnewcommand\System{\item[\algorithmicsystem]}

\algnewcommand\algorithmicnoise{\textbf{Noise}}
\algnewcommand\Noise{\item[\algorithmicnoise]}

\algnewcommand\algorithmicinitialization{\textbf{Initialization}}
\algnewcommand\Initialization{\item[\algorithmicinitialization]}

\algnewcommand\algorithmicfilter{\textbf{Filter}}
\algnewcommand\Filter{\item[\algorithmicfilter]}

\algrenewcommand\algorithmicindent{0.3cm}%

\makeatletter  
\def\colvec#1{\expandafter\colvec@i#1,,,,,,,,,\@nil}
\def\colvec@i#1,#2,#3,#4,#5,#6,#7,#8,#9\@nil{%
  \ifx$#2$ \begin{bmatrix}#1\end{bmatrix} \else
    \ifx$#3$ \begin{bmatrix}#1\\#2\end{bmatrix} \else
      \ifx$#4$ \begin{bmatrix}#1\\#2\\#3\end{bmatrix}\else
        \ifx$#5$ \begin{bmatrix}#1\\#2\\#3\\#4\end{bmatrix}\else
          \ifx$#6$ \begin{bmatrix}#1\\#2\\#3\\#4\\#5\end{bmatrix}\else
            \ifx$#7$ \begin{bmatrix}#1\\#2\\#3\\#4\\#5\\#6\end{bmatrix}\else
              \ifx$#8$ \begin{bmatrix}#1\\#2\\#3\\#4\\#5\\#6\\#7\end{bmatrix}\else
                 \PackageError{Column Vector}{The vector you tried to write is too big, use bmatrix instead}{Try using the bmatrix environment}
              \fi
            \fi
          \fi
        \fi
      \fi
    \fi
  \fi 
}  
\makeatother